\documentclass[reprint,amsmath,amssymb,aps,pra,superscriptaddress]{revtex4-2}
\usepackage[english]{babel}
\usepackage{mathrsfs,mathtools,amsthm,amsfonts}
\usepackage{bm}
\usepackage{bbm}
\usepackage{physics}

\usepackage{graphicx}
\usepackage[small,tight,FIGTOPCAP]{subfigure}

\usepackage{xcolor}
\usepackage{hyperref}
\hypersetup{colorlinks=true, citecolor=blue, linkcolor=black, urlcolor=blue}

\usepackage{enumerate}
\usepackage[top=1in,bottom=1in,left=1in,right=1in]{geometry}
\usepackage{hyperref, graphicx, caption, subcaption, float, wrapfig, pgfplots, booktabs, array}
\usepackage{tikz}
\usetikzlibrary{calc}
\usepackage[section]{placeins}
\usepackage{subcaption}
\pgfplotsset{compat=1.18} 

\newcommand{\R}{\mathbb{R}}

\renewcommand{\epsilon}{\varepsilon}

\DeclareMathOperator{\dom}{dom}

\newcommand{\Ac}{\mathcal{A}}

\newcolumntype{C}{>{\centering\arraybackslash}m{6em}}

\begin{document}
\title{\texorpdfstring{$\mathcal{PT}$}{PT}-symmetry enabled stable modes in multi-core fiber}
\author{Tamara Gratcheva}
\affiliation{F. Maseeh Department of Mathematics \& Statistics Portland State University, Portland, Oregon 97207}
\email{tgrat2@pdx.edu}
\author{Yogesh N. Joglekar}
\affiliation{Department of Physics, Indiana University Indianapolis (IUI), Indianapolis, Indiana 46202}
\author{Jay Gopalakrishnan}
\affiliation{F. Maseeh Department of Mathematics \& Statistics Portland State University, Portland, Oregon 97207}

\begin{abstract}
Open systems with balanced gain and loss, described by parity-time ($\mathcal{PT}$-symmetric) Hamiltonians have been deeply explored over the past decade. Most explorations are limited to finite discrete models (in real or reciprocal spaces) or continuum problems in one dimension. As a result, these models do not leverage the complexity and variability of two-dimensional continuum problems on a compact support. Here, we investigate eigenvalues of the Schr\"{o}dinger equation on a disk with zero 
boundary condition, in the presence of constant, $\mathcal{PT}$-symmetric, gain-loss potential that is confined to two mirror-symmetric disks. We find a rich variety of exceptional points, re-entrant $\mathcal{PT}$-symmetric phases, and a non-monotonic dependence of the $\mathcal{PT}$-symmetry breaking threshold on the system parameters. By comparing results of two model variations, we show that this simple model of a multi-core fiber supports propagating modes in the presence of gain and loss. 
\end{abstract}

\maketitle


\section{Introduction}
\label{sec:intro}
Over the past 25 years, research on non-Hermitian Hamiltonians with real spectra has burgeoned across disparate topics in physics, spanning mathematical physics~\cite{Bender2001,Mostafazadeh2002,Mostafazadeh2003-equivalence,Mostafazadeh2010}, optics and photonics~\cite{Ruter2010,Regensburger2012,Feng2017}, metamaterials~\cite{Sotiris2019}, acoustics~\cite{Zhu2014}, electrical circuits~\cite{Schindler2011,Chitsazi2017,LeonMontiel2018,Wang2020}, condensed matter physics~\cite{Lee2018,Ashida2020}, and open quantum systems~\cite{AshClerk2019,Klauck2019,Li2019,Naghiloo2019,Ding2021,Quinn2023}. It started with Bender and Boettcher's discovery~\cite{Bender1998} that the Schr\"{o}dinger eigenvalue problem for a non-relativistic particle on an infinite line with complex potentials $V(x)=V_\mathrm{R}(x)+iV_\mathrm{I}(x)$ has purely real spectrum that is bounded below. Similar results are obtained for non-relativistic particle on a line with compact support~\cite{Levai2000,Znojil2001,Joglekar2012bagchi}, discrete tight-binding models on finite or infinite lattices~\cite{Jin2009,Bendix2009,Joglekar2010,Joglekar2013}, and even minimal models with $2\times 2$ Hamiltonians. In each case, the non-Hermitian Hamiltonian $H$ --- a continuum, unbounded operator or a matrix --- is invariant under combined operations of parity $\mathcal{P}$ and time-reversal $\mathcal{T}$. This antilinear $\mathcal{PT}$-symmetry guarantees purely real or complex conjugate eigenvalues~\cite{Ruzicka2021}. 

After their experimental realizations in numerous platforms, it has become clear that $\mathcal{PT}$-symmetric Hamiltonians accurately model open systems with balanced, spatially separated gain $(V_\mathrm{I}>0)$ and loss ($V_\mathrm{I}<0$)~\cite{Bender2016}. Their standard phenomenology is as follows: starting from the Hermitian Hamiltonian $H_0$ with real spectrum and Dirac-orthogonal eigenfunctions, as the imaginary part of the potential $V_\mathrm{I}(x)$ is increased, two or more real eigenvalues undergo level attraction, become degenerate, and then develop into complex-conjugate pairs. This eigenvalue degeneracy, called exceptional point (EP) degeneracy~\cite{Kato1995,Miri2019,Ozdemir2019}, is characterized by the coalescence of corresponding eigenfunctions and lowering of the rank of the Hamiltonian operator. Due to the antilinearity of the $\mathcal{PT}$-operator, an eigenfunction $f_n(x)$ is simultaneously an eigenfunction of the $\mathcal{PT}$ operator with unit eigenvalue if and only if the corresponding eigenvalue $\lambda_n$ is real; if $\lambda_n$ is complex, then it follows that $\mathcal{PT}f_n(x)$ is an eigenfunction with complex-conjugate eigenvalue $\lambda_n^*$. The transition across the EP from a real spectrum to one with complex-conjugate eigenvalues is called $\mathcal{PT}$-symmetry breaking transition, since the corresponding eigenfunctions lose that symmetry, $\mathcal{PT}f_n(x)\neq f_n(x)$. 

Here, we investigate a two-dimensional continuum model on a compact domain subject to hard-wall (vanishing eigenfunctions) boundary condition in the presence of constant $\mathcal{PT}$-symmetric complex-valued potentials. In one dimension, such potential leads to a single $\mathcal{PT}$-symmetry breaking transition when the strength of the imaginary part of the potential, $\gamma,$ exceeds a threshold $\gamma_\mathrm{PT}$ set by the Hermitian Hamiltonian $H_0$. We will show that the two-dimensional case differs dramatically. It leads to multiple transitions where pairs of stable modes (real spectra) change into amplifying and leaky modes (complex conjugate eigenvalues) as $\gamma$ is increased. More surprisingly, we also find $\mathcal{PT}$-restoring transitions where, as the pure gain-loss potential $V_\mathrm{I}$ is increased, amplifying and leaky modes are pairwise stabilized. We argue that this unusual behavior arises due to complex interplay between the size of the modes in the Hermitian limit, and the size of the gain-loss region. 

The plan of the paper is as follows. In Sec.~\ref{sec:intro} we introduce the model and recall the Hermitian-limit results for a cylindrical waveguide. Section~\ref{sec:num} contains the outline of the numerical procedure we use for discretization. Results for eigenspectra and eigenfunctions across multiple $\mathcal{PT}$-breaking and restoring transitions are shown in Sec.~\ref{sec:results}. 
Section~\ref{sec:disc} concludes the paper. 


\section{\texorpdfstring{$\mathcal{PT}$}{PT}-symmetric fiber with circular cross-section}
\label{sec:model}
As a physical example, we consider a lengthwise uniform, multi-core fiber with circular cross-section of radius $R=1$ (purple) centered at the origin in the $x_1$-$x_2$ plane, a lossy core of radius $\rho$ centered at distance $d/2$ from the origin (green), and a gain-medium core of the same radius $\rho$ centered at the mirror-symmetric location (pink). The position-dependent index of refraction in the fiber is given by $n(x_1,x_2)=n_0+\delta n(x_1,x_2)$ where $n_0\sim 1$, and the index contrast $\delta n\sim 10^{-4}\ll n_0$~\cite{Harter2016,Harter2018}. Gain and loss can then be modeled by introducing negative and positive imaginary parts to the index contrast respectively, $\delta n=\delta n_R\mp i\delta n_I$. The Maxwell's equation for a transverse-magnetic (TM) mode, characterized by a vanishing electric field at the boundary, implies that the field ${\bf E}(x)=E(x_1,x_2)\exp[i(k_3x_3-\omega t)]\hat{x}_3$, propagating along the fiber, is given by 
\begin{align}
    \label{eq:me}
    -\left[\Delta'+\frac{2n_0\omega^2}{c^2}\delta n\right]E(x)=\left[\frac{n_0^2\omega^2}{c^2}-k_3^2\right]E(x)
\end{align}
where $\Delta'\equiv(\partial^2_{x_1}+\partial^2_{x_2})$ is the in-plane Laplacian with dimensions of inverse-area and $c$ is the speed of light in vacuum. After suitable rescaling, Eq.(\ref{eq:me}) can be mapped onto a Schr\"{o}dinger-like eigenvalue problem~\cite{Christodoulides2003,Szameit2010}, 
\begin{align}
\label{eq:se1}
& -\Delta f_n(x)+ V(x)f_n(x)=\lambda_n f_n(x),\\
\label{eq:se2}
& V(x)=V_B-\frac{2 n_0 R^2\omega^2}{c^2}\delta n(x_1,x_2),\\
\label{eq:se3}
&\lambda_n= V_B+\frac{n_0^2 R^2\omega^2}{c^2}-R^2k_3^2.
\end{align}
Here, $\Delta=R^2\Delta'$ is the dimensionless Laplacian, $\lambda_n$ denotes the dimensionless eigenvalue, $f_n$ is the corresponding eigenmode, and $V_B$ sets the zero for the dimensionless potential $V(x)$. Changing $V_B$ shifts the overall spectrum but does not change the level differences $\Delta\lambda_{mn}\equiv \lambda_m-\lambda_n$. We see from Eq.(\ref{eq:se2}) that a positive index-contrast $\delta n>0$ acts as an attractive potential for the electric field. We proceed to identify $V(x)$ for the specific geometry we will consider.

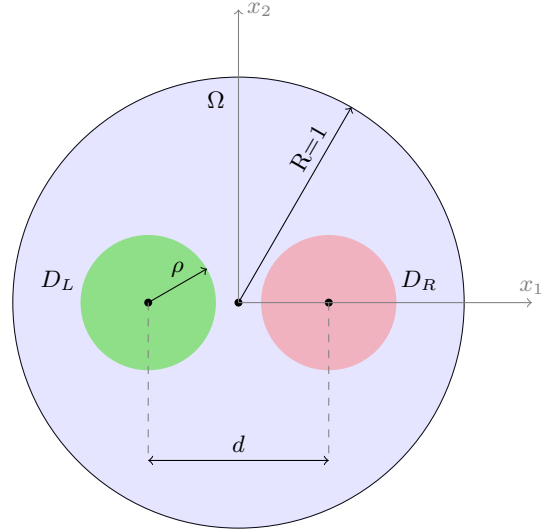
\begin{figure}[H]
  \centering
    \begin{tikzpicture}[scale=3]    

      \coordinate (O) at (0, 0);
      \coordinate (CL) at (-0.4, 0);
      \coordinate (CR) at (0.4, 0);
      \draw (O) circle (1);
      \fill[blue!50, opacity=0.2] (O) circle (1);
      \fill[green!70!brown, opacity=0.5] (CL) circle (0.3); 
      \fill[red!50, opacity=0.5] (CR) circle (0.3); 
      
      \draw[->] (O)--(60:1) node [near end, sloped, above] {R=1};
      \fill (O) circle (0.5pt);
      \fill (CL) circle (0.5pt);
      \fill (CR) circle (0.5pt);    
      \draw[->] (CL)--++(30:0.3) node[midway, above] {$\rho$};    

      \draw[->, gray] (O)--++(1.3, 0) node[above] {$x_1$};
      \draw[->, gray] (O)--++(0, 1.3) node[right] {$x_2$};

      \coordinate (CLd) at ($(CL)+(0, -0.7)$);
      \coordinate (CRd) at ($(CR)+(0, -0.7)$);
      

      \draw[dashed, gray] (CL)--(CLd);
      \draw[dashed, gray] (CR)--(CRd);    
      \draw[<->] (CLd)--(CRd) node[midway, above] {$d$};

      \node at ($(CL)-(0.4,-0.1)$) {$D_L$};
      \node at ($(CR)+(0.4,+0.1)$) {$D_R$};
      \node at (-0.1, .9) {$\Omega$};
      
    \end{tikzpicture}
    \caption{Schematic cross-section of cylindrical, multi-core fiber. The radius $R=1$ sets the length-scale. The pink region $D_R$, centered at $x_1=d/2$ with radius $\rho$ denotes the gain region, and the green region $D_L$, centered at mirror-symmetric point $x_1=-d/2$ with the same radius denotes the loss region. When the gain-loss regions have no real-part for the index-contrast with the rest of the fiber (purple), i.e. $V_0=0$, the eigenmodes are not just confined to the regions $D_L$ and $D_R$.}
    \label{fig:geom}    
\end{figure}


Let $\mathcal{B}_r(x^0_1, x^0_2) = \{ (x_1, x_2) \in \R^2:  (x_1 - x^0_1)^2 + (x_2 - x^0_2)^2  < r^2 \}$ denote the disk of dimensionless radius $r$ centered at $x^0$. Our problem is set in the domain $\Omega = \mathcal{B}_1(0, 0)$ where a purely imaginary gain-loss potential is introduced into non-intersecting left and right subdomains $D_L = \mathcal{B}_\rho(-d/2, 0)$ and $D_R = \mathcal{B}_\rho(d/2, 0)$ where $\rho<R$ and $2\rho\leq d\leq 2(R-\rho)$ ensures that the two domains do not intersect (Fig.~\ref{fig:geom}).
The eigenvalue problem is to find complex-valued, square-integrable functions on $\Omega$, $f_n(x_1,x_2)\in L^2(\Omega),$ that are in the domain of the operator $\Ac$ defined below, and that vanish on the boundary $\partial\Omega$, together with  complex numbers $\lambda_n$, such that
\begin{align}
\label{eq:schr}
\Ac f_n(x) &=(-\Delta+V)f_n(x)=\lambda_n f_n(x) && \text{in } \Omega,    
\end{align}
with 
\begin{align}
\label{eq:v}
  V(x_1, x_2) =  V_B + \begin{cases}
        V_0-i\gamma & \text{ if } (x_1, x_2) \in D_L,\\
        V_0+i\gamma & \text{ if } (x_1, x_2) \in D_R, \\
                   0   &  \text { otherwise. }
        \end{cases}  
\end{align}
We set the background potential $V_B=1$. It is constant over the entire domain $\Omega$. It does not affect the eigenvalue {\it differences} and the resulting $\mathcal{PT}$-symmetry breaking threshold where the spectrum transitions between purely real and complex-conjugate pairs. We define the parity operator $\mathcal{P}:L^2(\Omega) \to L^2(\Omega)$ by $(\mathcal{P}f)(x_1, x_2)=f(-x_1, x_2)$, i.e., $\mathcal{P}$ mirrors functions about the second axis. It is easy to see that $\mathcal{P}$ is a linear, self-adjoint, and unitary operator in $L^2(\Omega)$. The antilinear time-reversal operator $\mathcal{T}: L^2(\Omega)\to L^2(\Omega)$ is given by $(T f)=f^*$. An operator $H$ is called $\mathcal{PT}$-symmetric if it commutes with the antilinear operator $\mathcal{PT}$, 
\begin{align}
\label{eq:pt}
    \mathcal{PT}H=H\mathcal{PT}.
\end{align}
For unbounded operators $H$ defined on a proper subspace $\dom(H) \subset L^2(\Omega)$ rather than all of $L^2(\Omega)$, namely
$H: \dom(H) \to L^2(\Omega)$, Eq.\eqref{eq:pt} means $f,\mathcal{PT}f\in\dom H$ and the equality \eqref{eq:pt} holds. The operator of interest to us, $\mathcal{A}=-\Delta+V(x_1,x_2)$, is unbounded, and its domain is given by $\dom(\Ac)= H^2(\Omega )\cap \mathring{H}^1(\Omega)$. Here $H^k(\Omega)$ denotes the Sobolev space of square-integrable functions all of whose derivatives of order at most $k \ge 1$ are also square integrable and $\mathring{H}^1(\Omega)$ denotes the subspace of $H^1(\Omega)$-functions that vanish on the boundary~$\partial \Omega$. It is straightforward to check that $\Ac$ is $\mathcal{PT}$-symmetric. 

Note that when the index contrast $V_0<0$ is sufficiently large in magnitude, the modes $f_n(x_1,x_2)$ become largely confined to the gain and loss regions. Then our problem reduces to the well-studied $\mathcal{PT}$-symmetric coupler where the size of the mode is comparable to the size of the gain-loss region, the system can be effectively modeled by a $2\times 2$ Hamiltonian, and undergoes a single $\mathcal{PT}$-symmetry breaking transition~\cite{Ruter2010,ElGanainy2018}. Instead, we choose
$V_0=0$ to ensure the eigenfunctions $f_n(x_1,x_2)$ are spread over the entire disk $\Omega$. 

To investigate the eigenvalues of $\Ac(\gamma)$, we start with the Hermitian limit of Eq.\eqref{eq:v}, $\gamma=0$. In this case, the cylindrical symmetry in the $x_1$-$x_2$ plane gives unnormalized eigenfunctions in polar coordinates $r, \theta$, 
\begin{equation}
  \label{eq:h0eig}
  \begin{gathered}
    f^{0}_{\pm mp}(r,\theta)=J_m(r\sqrt{\lambda_{mp}-V_B})
    e^{\pm im\theta}, 
\end{gathered}  
\end{equation}
for $m \ge 0$ and $p \ge 1$,  where the corresponding eigenvalue $\lambda_{mp}$ is determined by the $p^\text{th}$ zero of the $m^\text{th}$ Bessel function, $J_m(\sqrt{\lambda_{mp}-V_B})=0$, which enforces the hard-wall boundary condition $f|_{\partial\Omega}=0$. Except for $m=0$, these solutions with $\exp(\pm im\theta)$ are degenerate, and represent positive and negative angular momentum states respectively. The semi-analytically obtained eigenvalues $\lambda_{mp}$ of $\Ac(\gamma=0)$ are the starting point for computing eigenvalue trajectories $\lambda_{mp}(\gamma)$. They also serve to verify our numerical methodology by benchmarking it against the $\gamma=0$ case. 


\section{Numerical discretization in arbitrary 2D domains}
\label{sec:num}

In a one-dimensional interval, the discretization of the Schr\"{o}dinger operator with hard-wall boundary condition leads to a tridiagonal matrix with no corner elements, whose absence enforces the boundary conditions. Two-dimensional domains, on the other hand, require more care. Let us denote the complex $L^2(\Omega)$-inner product by $\braket{\cdot}$. For any smooth function $g$ vanishing on $\partial\Omega$, the eigenvalue equation~\eqref{eq:schr} implies
\begin{align*}
\bra{g}\ket{Af_n}=\bra{\nabla g(x)}\ket{\nabla f_n(x)}+\bra{g(x)}\ket{V(x)f_n(x)}.
\end{align*}
The finite element method imposes the same equation on the {Lagrange finite-element}~\cite{ErnGuerm21} space $X_h$ consisting of continuous functions, vanishing on the boundary $\partial \Omega$, which are polynomials of degree at most $p$ in each mesh element; in our computations,
we use $p=5.$ 
Here the mesh is a geometrically conforming mesh of triangles subdividing the domain, respecting the material interfaces, with curved elements with higher density near the circular boundaries and interfaces. The subscript $h$ indicates the maximal diameter of all elements in the mesh. As $h$ becomes smaller or $p$ becomes larger, the discretization becomes finer and $\dim X_h$ becomes larger.

Our numerical method computes the eigenvalues of a discretization $\Ac_h: X_h \to X_h$ of the infinite-dimensional operator $\Ac$. It is defined by
\begin{align}
    \label{eq:Ach-defn}
  \bra{g_h}\ket{\Ac_hf_h}=\bra{\nabla g_h}\ket{\nabla f_h}+\bra{g_h}\ket{Vf_h}
\end{align}
for all $f_h,g_h\in X_h$. Namely, we compute an eigenvalue approximation~$\lambda_{h,n}$ and right eigenfunction $f_{h,n}$ satisfying
\begin{align}
    \label{eq:Ah-epair}
  \Ac_h f_{h,n} = \lambda_{h,n}f_{h,n}.
\end{align}
Standard finite-element theory~\cite{BabusOsbor91} can be used to show that the approximate eigenpairs $(f_{h,n},\lambda_{h,n})$ converge to the exact ones under suitable assumptions as $h \to 0$; the symmetry of the mesh is immaterial in obtaining such convergence. The right eigenfunction $f_h\in X_h$ in~\eqref{eq:Ah-epair} is equivalently given by 
\begin{equation}
  \label{eq:fem-form}
  \bra{g_h}\ket{\Ac_h f_h}=\lambda_h\bra{g_h}\ket{f_h} \qquad \text{ for all } g_h \in X_h.  
\end{equation}
Using a non-orthogonal basis $\psi_i$ of finite-element shape functions, Eq.~\eqref{eq:fem-form} can be converted to a matrix eigenvalue problem
\begin{equation}
\label{eq:abeigenvalue}
   A x = \lambda B x 
\end{equation}
where $A_{ij} = \bra{\psi_i}\ket{\Ac_h \psi_j}$ and $B_{ij} =\bra{\psi_i}\ket{\psi_j}$. This generalized eigenproblem is then solved for a cluster of selected eigenvalues using a contour integral method called the FEAST algorithm~\cite{Poliz09, GopalGrubiOvall20a}, which can also compute the
corresponding eigenmodes for the nonselfadjoint eigenproblem~\cite[Algorithm~1]{GopalParkeVande22}.
The size of the eigenproblem for each $\gamma$ value, namely 
$\dim X_h$, is determined by the degree $p$, the geometrical parameters ($\rho$ and $d$) and how it  constrains the mesh size $h$; in our computations  $\dim X_h$ ranged
from  8000 to 16000.

While much of our ensuing analysis use meshes without symmetry, we have also experimented with meshes with parity symmetry that are invariant under reflection by the vertical axis ($x_1=0$). On such meshes, the discretized $\Ac_h$ is exactly $PT$-symmetric; specifically, \eqref{eq:Ach-defn} implies that
\begin{equation}
  \label{eq:PTA-h}
    \mathcal{PT}\Ac_h f_h = \Ac_h\mathcal{PT}f_h
\end{equation}
for all $f_h \in X_h$, recovering the perfect analogue of Eq.\eqref{eq:pt} on the discrete space $X_h$. In practice, this implies that exactly real eigenvalues
are recovered with imaginary parts of the order of machine precision when using
meshes with parity symmetry. In contrast, when using meshes without the symmetry, the same eigenvalues are approximated by numbers whose 
the imaginary parts are generally not machine zero, but rather 
approach zero up to discretization errors.
Also  note that since eigenfunctions are defined only up to a scaling factor, 
the corresponding eigenmode intensities $|f_h|^2$ are also only defined up to a scaling factor. Hence we report intensities without explicitly
showing a color legend, with a blue-to-red colormap where blue denotes zero and red
denotes the maximum intensity value. 

Since the finite element discretization and the FEAST eigensolver do not depend on the shapes of fiber cross-section, gain domain, or the loss domain, this approach is uniquely suited to investigate the interplay among Hermitian mode structure, gain-loss geometry, and the widely tunable effective coupling between the gain and loss domains. 


\section{Numerical Results}
\label{sec:results}

We start with the typical results for the flow of lowest few eigenvalues $\lambda_{mp}(\gamma)$. Recall that at $\gamma=0$, all eigenvalues except the lowest one, $m=0$, are doubly degenerate. However, our judicious choice of the gain-loss domains ensures that there are no matrix elements for $V(x_1,x_2)$ between states $\pm m$ and therefore the spectrum $\lambda_n(\gamma)$ does not become immediately complex. 

\begin{figure}[H]   
\centering
\includegraphics[width=1.03\columnwidth]{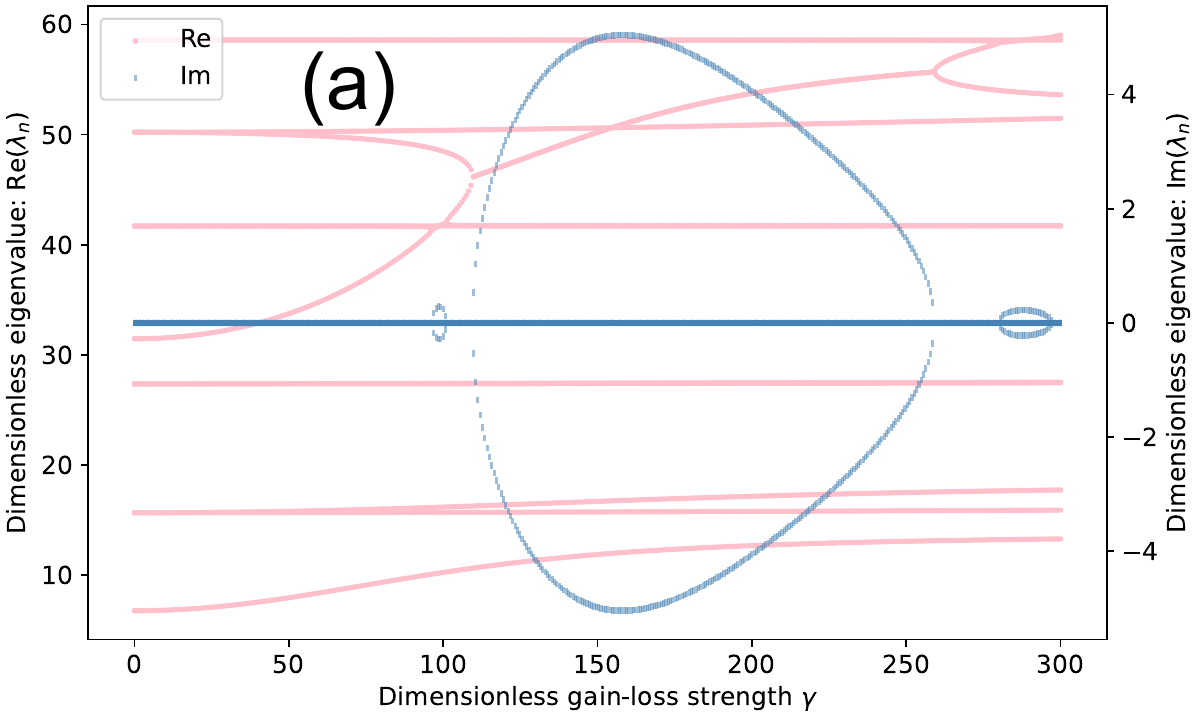} 
\includegraphics[width=0.49\columnwidth]{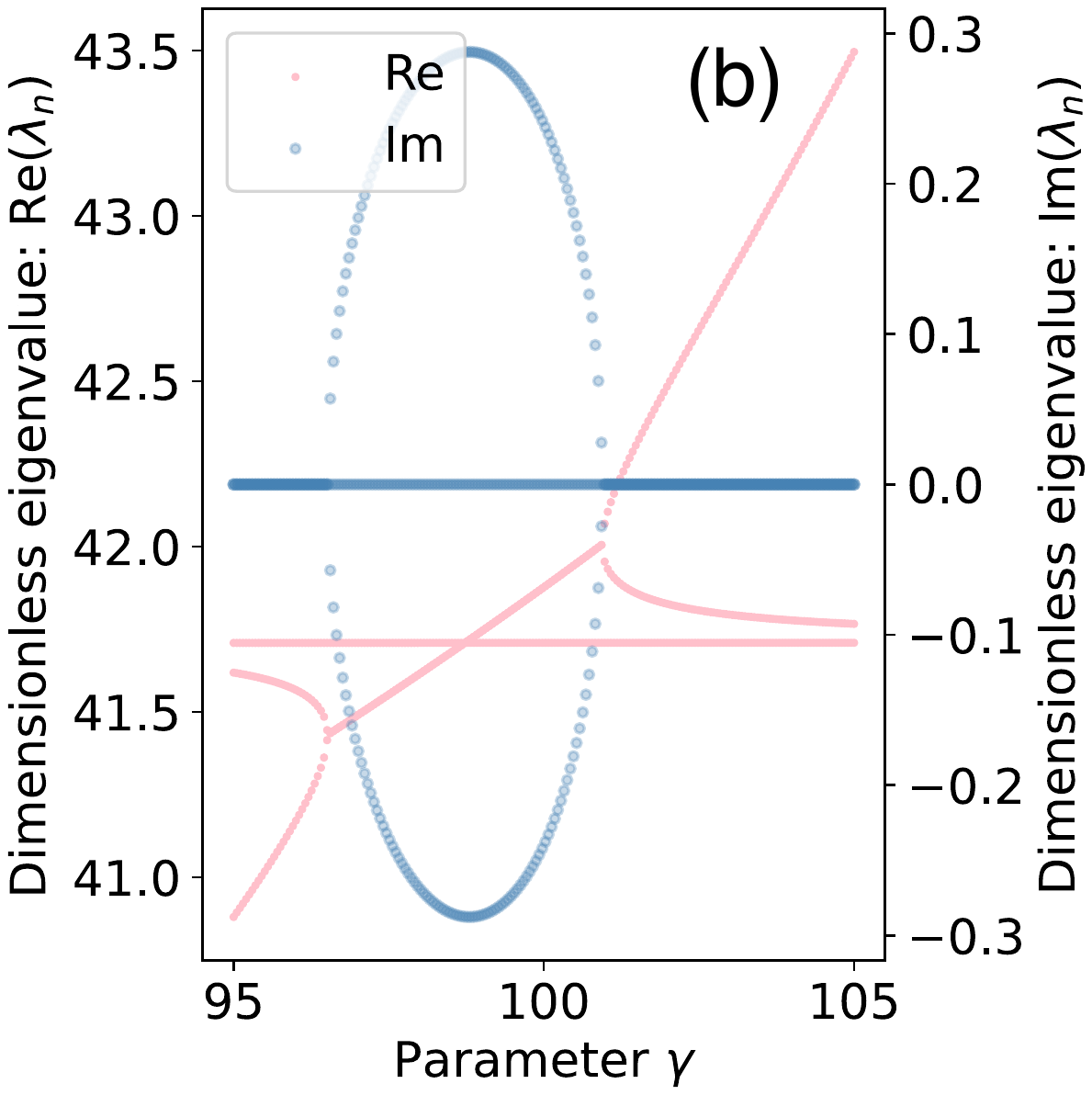}
\includegraphics[width=0.49\columnwidth]{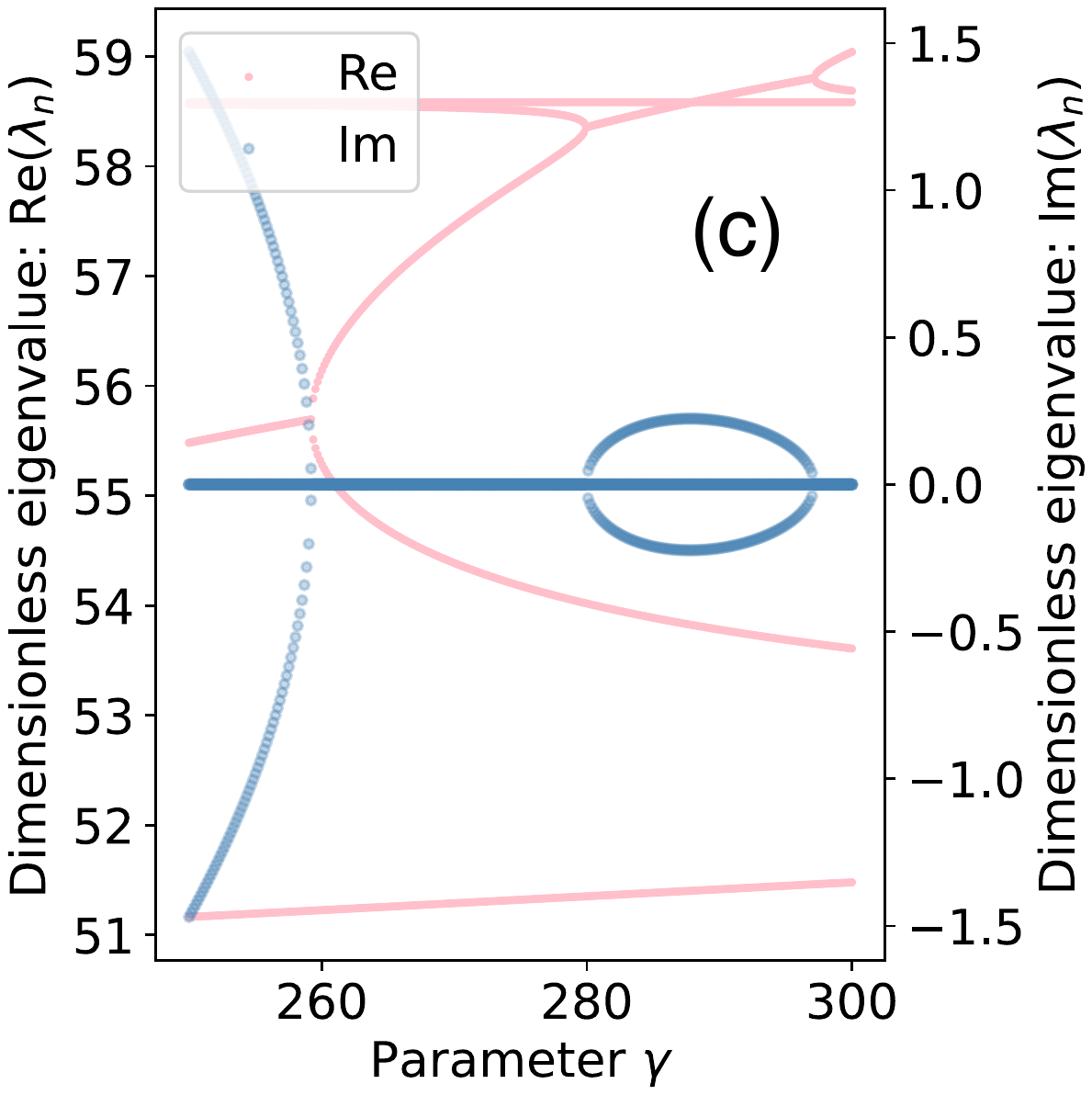}
\caption{Flow of eigenvalues $\lambda_{mp}(\gamma)$ for the first seven eigenvalues. 
All of them except $m=0$ cases are doubly degenerate at $\gamma=0$ and this degeneracy is lifted with increasing $\gamma$. The first $\mathcal{PT}$-symmetry breaking transition occurs at $\gamma=\gamma_\mathrm{PT}\approx 97$ immediately followed by $\mathcal{PT}$-restoring transition near $\gamma_\mathrm{PT}=102$ (detailed in the second plot). This is followed by a significantly broad $\mathcal{PT}$-broken region, and another small $\mathcal{PT}$-breaking and restoring transition. This re-entrant $\mathcal{PT}$-symmetric phase in a model with single gain-loss parameter is uncommon. These results are independent of the background potential value $V_B=1$, changing which uniformly shifts all eigenvalues $\lambda_n$ while leaving the flow-diagram unchanged.}
\label{fig:ptbreaking}
\end{figure}

In Fig.~\ref{fig:ptbreaking}a we track the real (pink) and imaginary (blue) parts of the lowest seven dimensionless eigenvalues as a function of dimensionless gain-loss strength $\gamma$ for a geometry with $d/R=0.3$ and $\rho/R=0.1$. As $\gamma$ is increased, we see level attraction, leading to degeneracy and emergence of complex conjugate pair, indicated by equal and opposite imaginary parts. The first such transition occurs near $\gamma=97$, shown in detail in the second plot, Fig.~\ref{fig:ptbreaking}b (with rescaled axes), and is followed by a $\mathcal{PT}$-restoring transition where the spectrum becomes purely real again near $\gamma=102$. It is followed by a large $\mathcal{PT}$-symmetry broken region in the range $110\leq\gamma\leq 265$, followed by another $\mathcal{PT}$-symmetric region. Figure~\ref{fig:ptbreaking}c shows the zoomed-in and rescaled-view of another such small window near $\gamma=290$.  

\begin{figure}[H]
 \centering
 \includegraphics[width=0.45\columnwidth]{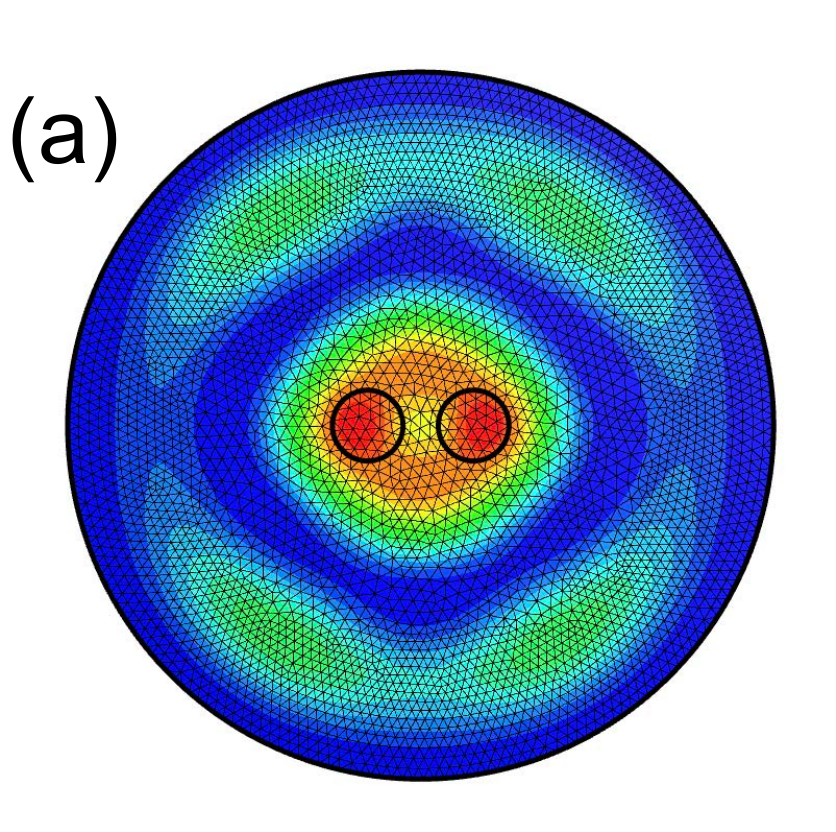}
 \hspace{1em}
 \includegraphics[width=0.45\columnwidth]{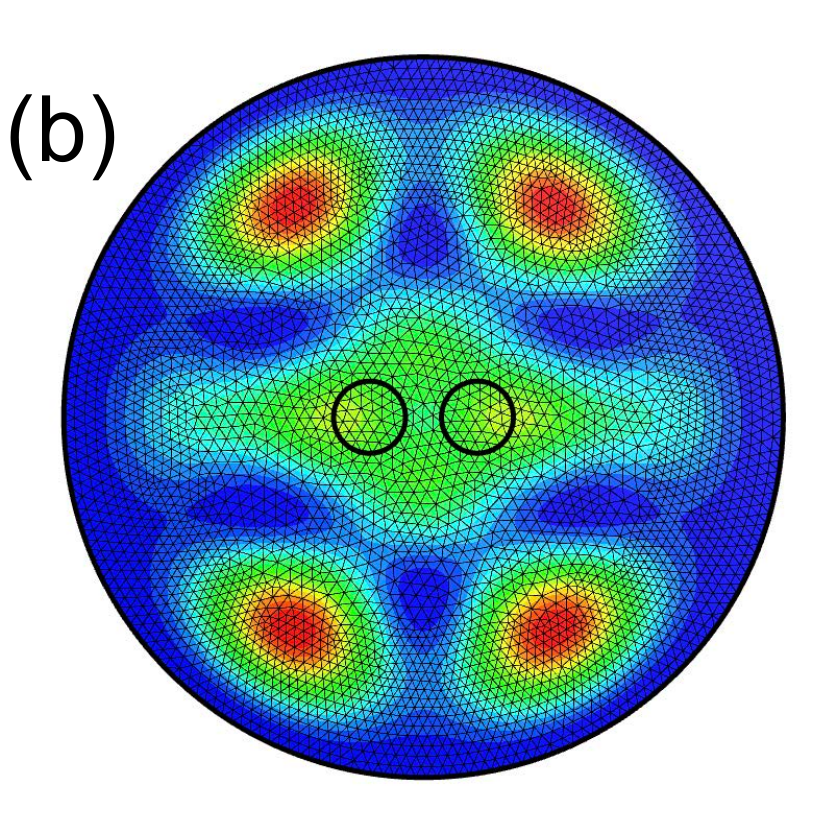}
 \includegraphics[width=0.45\columnwidth]{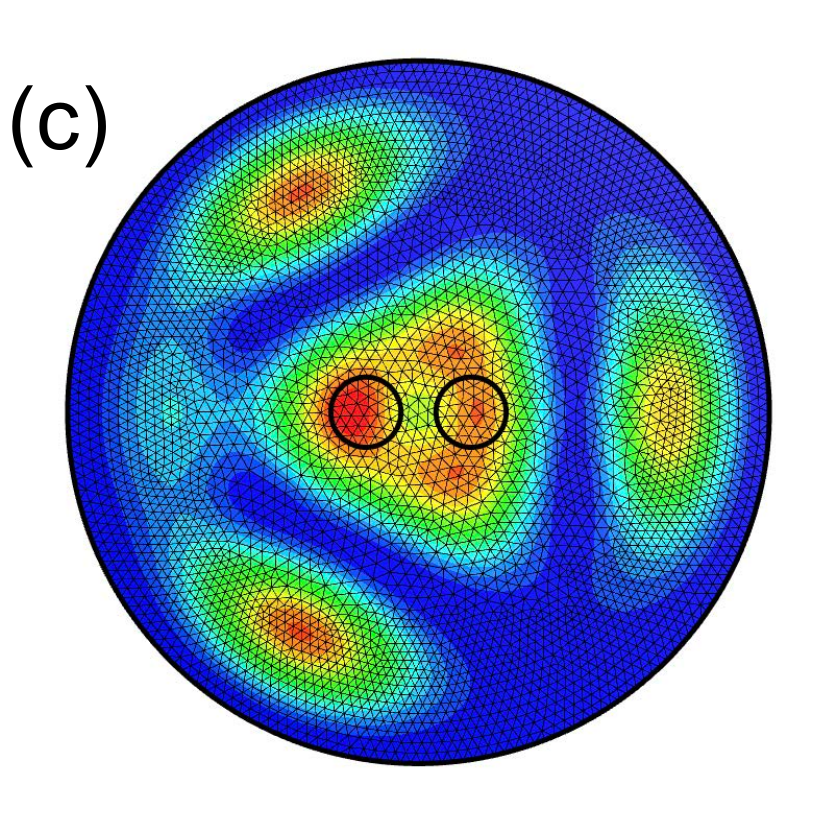}
  \hspace{1em}
 \includegraphics[width=0.45\columnwidth]{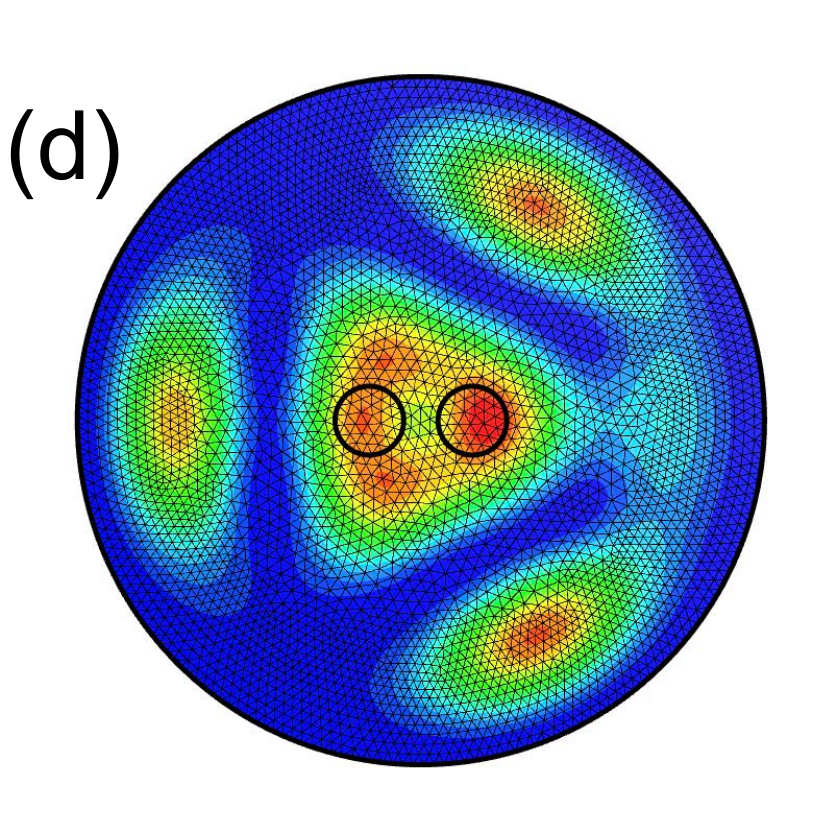}
 \includegraphics[width=0.49\columnwidth]{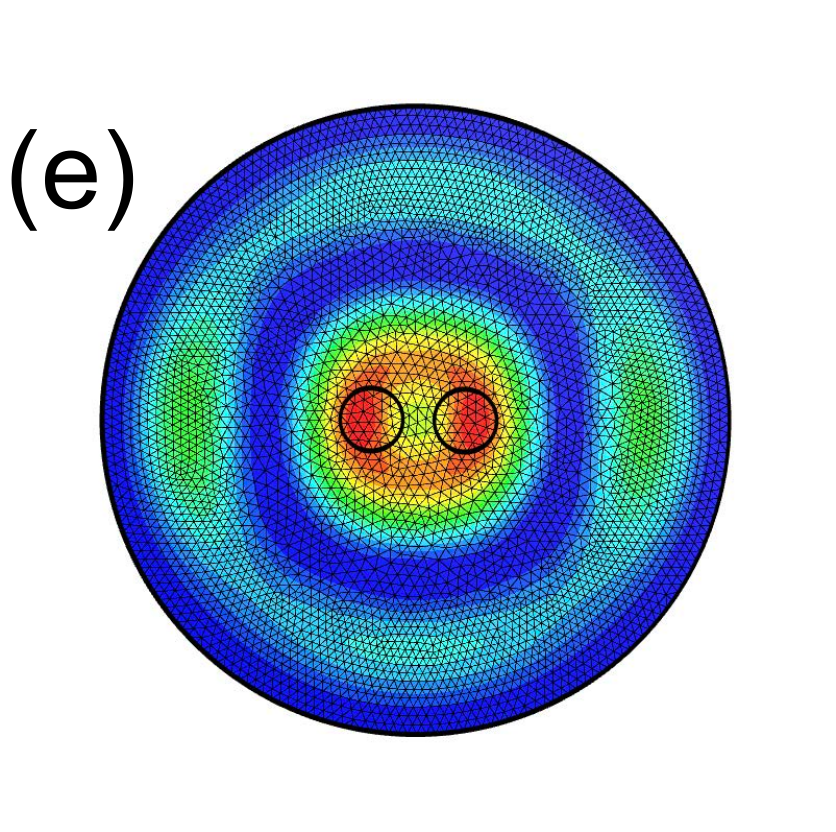}
 \includegraphics[width=0.49\columnwidth]{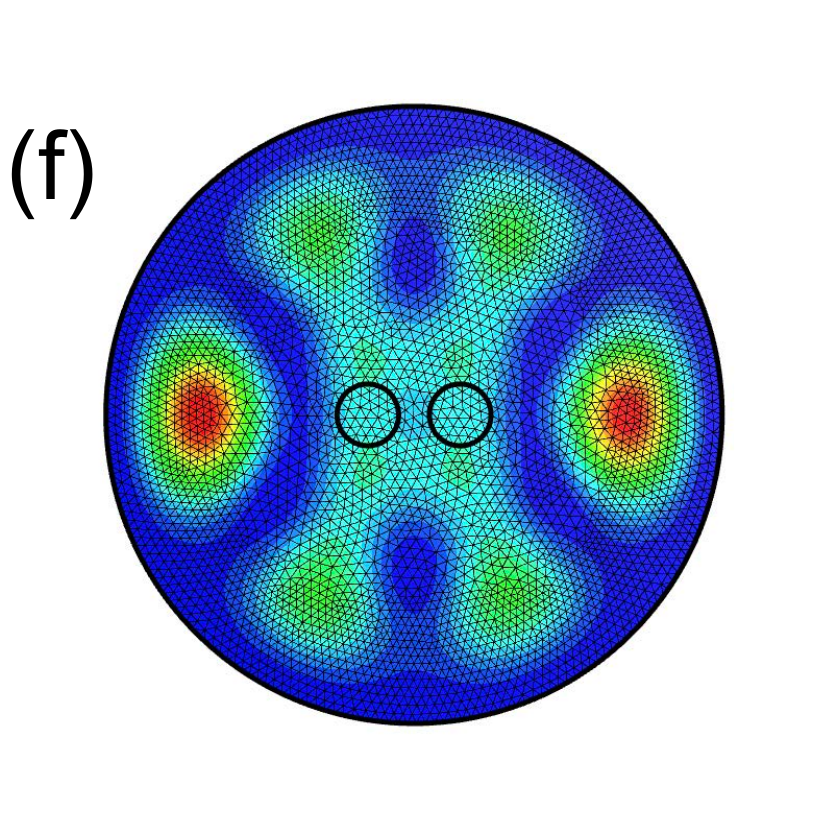} 
 \caption{Mode-intensity evolution for pair of eigenvalues in Fig.~\ref{fig:ptbreaking}b that become complex and then real again. The black circles denote the gain-loss regions with $d=0.3$ and $\rho=0.1$. Mode intensities in (a)-(b), at $\gamma=96$, are $\mathcal{PT}$-symmetric. Modes in (c)-(d),  at 
   $\gamma=99$, are in the $\mathcal{PT}$-symmetry-broken region:  the intensities of the two modes are mirror-images of each other, while each intensity, by itself, shows a broken $\mathcal{PT}$-symmetry. In (e)-(f),  gain-loss strength is increased further to $\gamma=102$, the eigenvalues become real again, leading to mode intensities that are individually mirror symmetric. 
 }
 \label{fig:modeintensities}
\end{figure}

The surprising emergence of multiple $\mathcal{PT}$-symmetry breaking transitions induced by variations of a single parameter $\gamma$ reflects the two dimensional nature of the underlying model. One-dimensional lattice or continuum models require potentials with different spatial ranges or functional forms for re-entrant $\mathcal{PT}$-symmetric phases to arise~\cite{Joglekar2012bagchi,Liang2014,Agarwal2021}. It is also worth noting that, in contrast to traditional models, the lowest few eigenvalues continue to remain real.

In Fig.~\ref{fig:modeintensities} we show the evolution of the mode intensities with $\gamma$ for the pair of eigenvalues that become complex and then again real, see Fig.~\ref{fig:ptbreaking}b. These results are for $d/R=0.3$ and $\rho/R=0.1$, Each panel shows the triangular mesh, the gain-loss domains (black circles), and mode intensities. 
When the spectrum of the system is purely real at $\gamma=96$, (a)-(b), the intensities have equal weights on mirror-symmetric locations $x_1\leftrightarrow -x_1$. When the spectrum changes into complex-conjugate pair, the modes are preferentially localized in the loss region, (c), or its mirror-symmetric gain region, (d). We also note that although the $x_1\leftrightarrow -x_1$ symmetry is broken, the mirror-symmetry about the horizontal axis, $x_2\leftarrow -x_2$ continues to be obeyed by all eigenfunctions. When the gain-loss strength is further increased to $\gamma=102$, the spectrum becomes real again, and as shown in (e)-(f), the eigenmodes have equal weights in the gain and the loss regions. These typical results show that irrespective of the symmetry of the underlying mesh used for discretization, the numerically obtained eigenmodes also clearly show the $\mathcal{PT}$-symmetry breaking and restoring transitions.

Although we have shown results only for a single parameter set, Fig.~\ref{fig:ptbreaking}, the re-entrant $\mathcal{PT}$-symmetric phase occurs generically over a wide range of gain-loss domain sizes and separations. It is also important to note that when the index-contrast is increased, $|V_0|\gg 1$, $\mathcal{PT}$-symmetry breaking occurs via hybridization of the lowest $m=0,\pm 1$ modes, and the re-entrant $\mathcal{PT}$-symmetric phase disappears. These results suggest that the large spatial extent of the modes relative to the size of the gain-loss domains plays an important part.

\begin{figure}[H]
    \centering
    \includegraphics[width=0.49\textwidth]{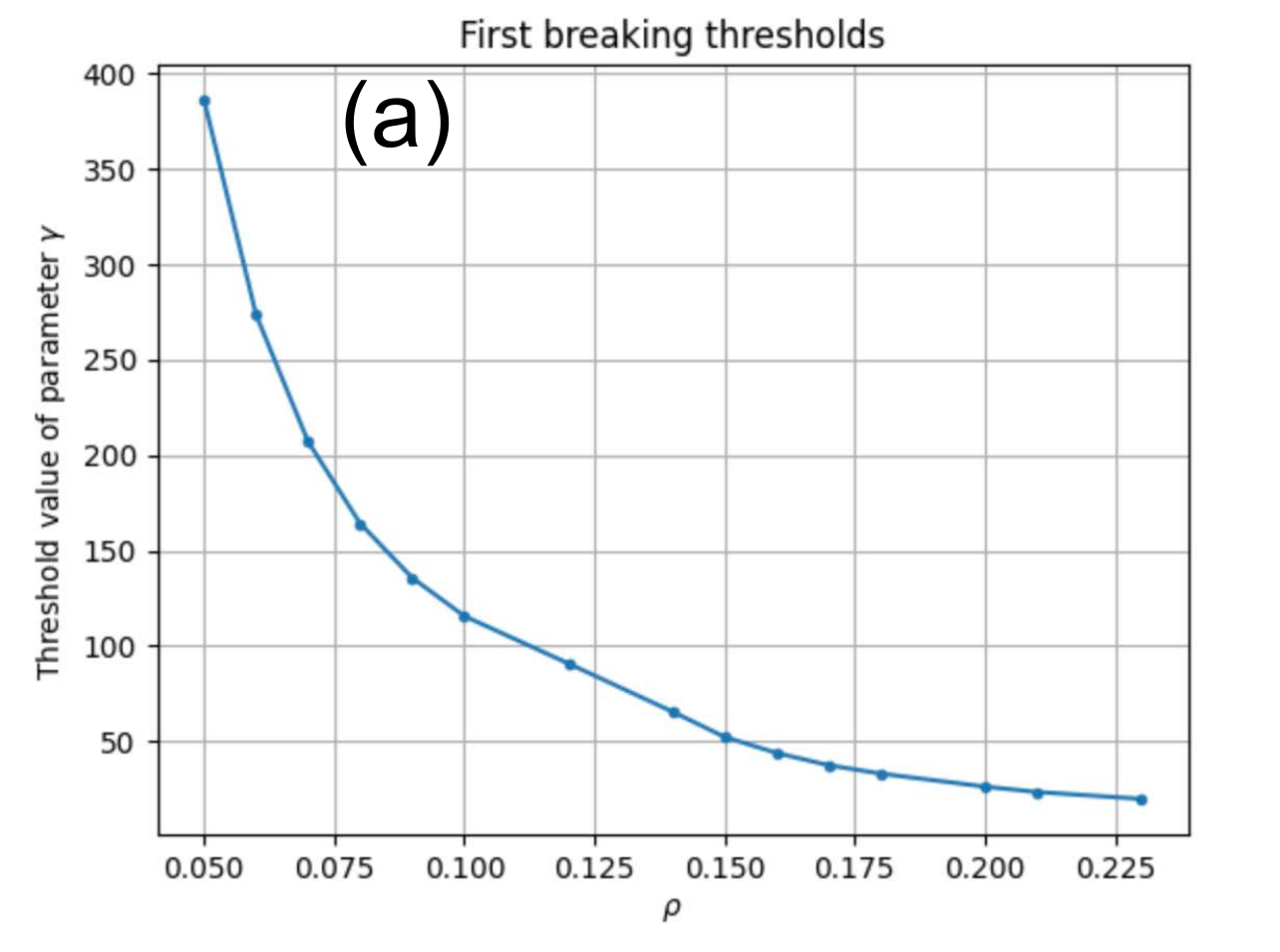}
    \includegraphics[width=0.49\textwidth]{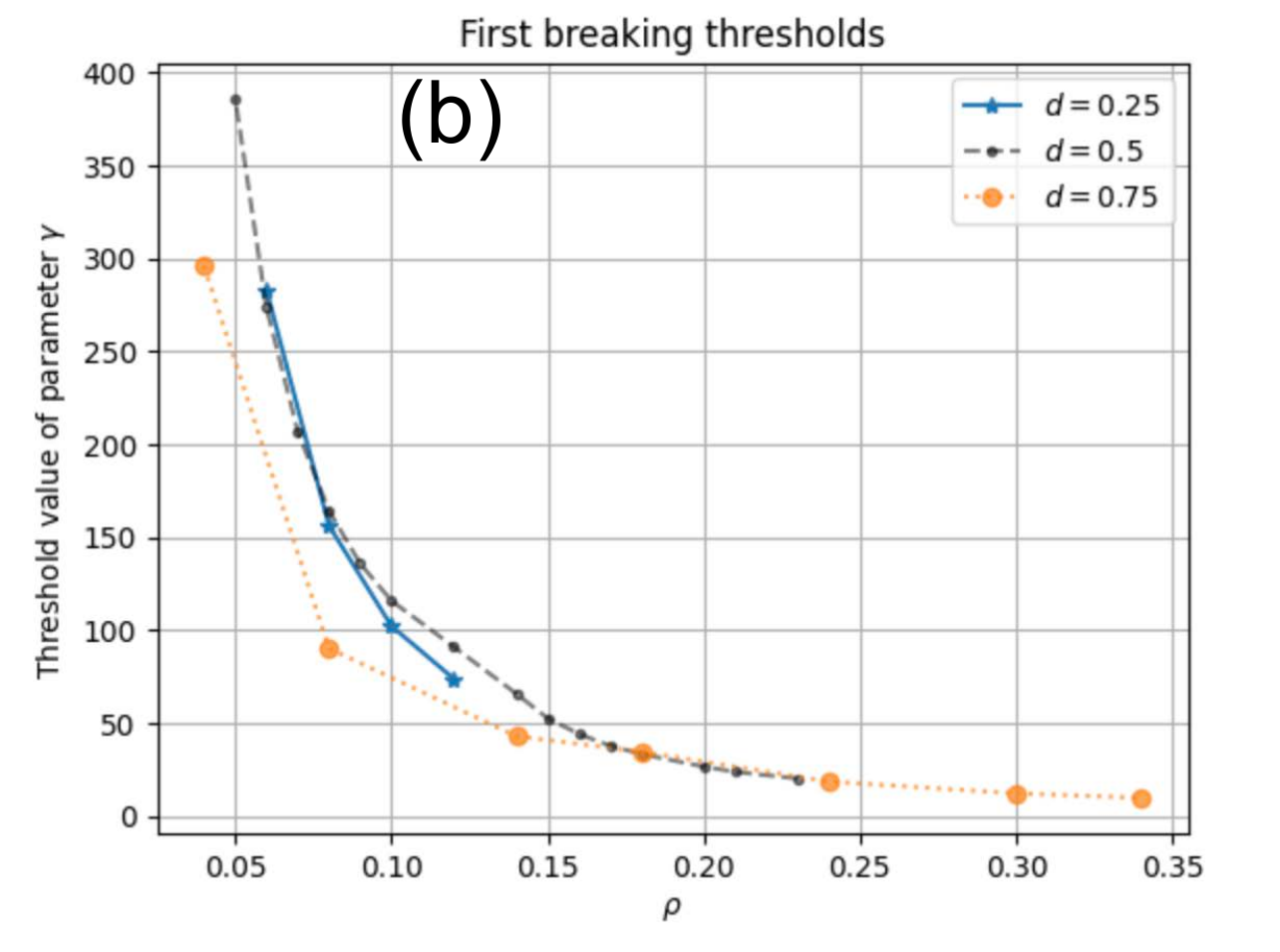}
    \caption{Dependence of $\gamma_\mathrm{PT}$, where first complex-conjugate eigenvalues emerge, on the dimensionless radius $\rho$ of the gain-loss domain. (a)~At distance $d/R=0.5$, $\gamma_\mathrm{PT}$ varies inversely with the $\rho\leq 2d$. (b)~This inverse behavior, expected from the effective-strength model with $\delta$-function gain-loss points, is valid for different values of $d$.}
    \label{fig:gammaptrho}
\end{figure}

Next, we investigate the dependence of the $\mathcal{PT}$-threshold strength $\gamma_\mathrm{PT}$ on the radius $\rho$ of the gain-loss domains $D_L,D_R$ and the center-to-center distance $d$ between them, using the results shown in Fig.~\ref{fig:gammaptrho}. 
Plot~(a) shows that $\gamma_\mathrm{PT}$ varies inversely with $\rho$ at $d/R=0.5$. This is expected because the  "effective gain-loss strength"  is given by $\gamma\pi\rho^2$. In the limit when $\rho/d\ll 1$, the system goes over to two, localized $\delta$-function-like gain and loss potentials~\cite{Joglekar2010a,Joglekar2012bagchi} with a finite threshold that depends on this effective strength. This inverse-relationship is valid for general $d$, Fig.~\ref{fig:gammaptrho}b over the possible range of $\rho< 2d$.

\begin{figure}[H]
    \centering
    \includegraphics[width=0.49\textwidth]{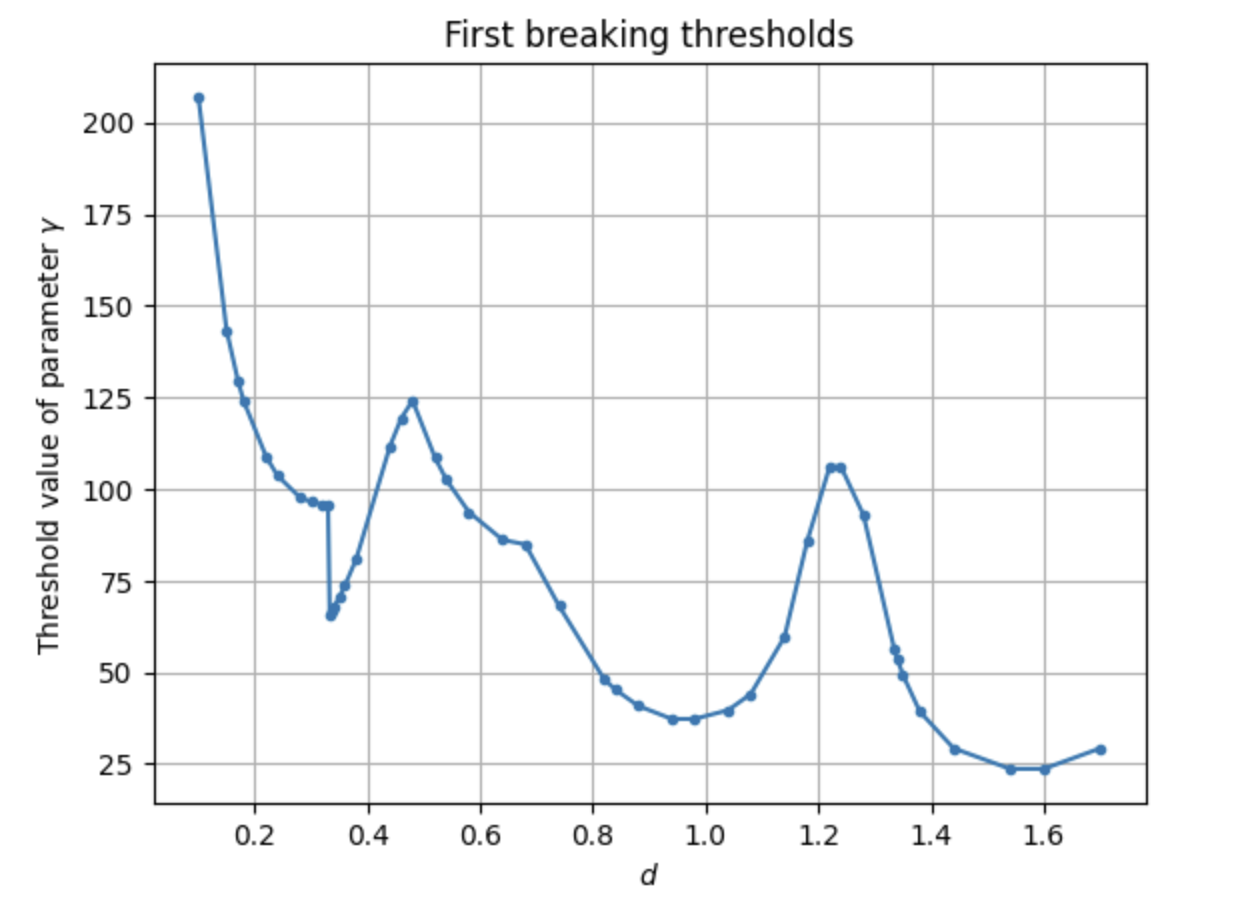}
    \caption{Dependence of the first threshold $\gamma_\mathrm{PT}$ on the distance $d$ for a fixed radius 
      $\rho=0.1$
      of the gain-loss domains. After an initial decay, the threshold is recovered even as the gain and loss regions move farther away from each other. 
    }
    \label{fig:gammaptvsd}
\end{figure}

Lastly, in Fig.~\ref{fig:gammaptvsd} we show the variation of $\gamma_\mathrm{PT}$ with the center-to-center distance $d$. Recall that in traditional $\mathcal{PT}$-dimer models, where the modes are confined to the gain-loss regions, as the distance $d$ between the gain and loss regions increases, the effective coupling between them and subsequently the $\mathcal{PT}$-breaking threshold decreases. Here, however, we see that after an initial decay, $\gamma_\mathrm{PT}$ shows non-monotonic behavior. In particular, the threshold $\gamma_\mathrm{PT}$ is recovered even as the distance is increased six-fold from $d\sim 0.2$ to $d\sim 1.2$. This increase is due to competing effects of boundary proximity for the gain and loss regions, and increased distance between them~\cite{Joglekar2010,Joglekar2012bagchi}. Note that for small values of $d$ 
in Fig.~\ref{fig:gammaptvsd}, when $d < 2 \rho$,  the gain and loss disks overlap; within the overlap, since gain and the loss cancel each other out, the values of $V$ are real. Then the breaking thresholds are high, being determined by  small nonoverlapping slivers of gain and loss. As $d$ is increased, the overlap decreases (with no overlap when $d>2\rho$), thus explaining the initial decay of the breaking threshold values.

\section{Conclusion}
\label{sec:disc}
In this work, we have numerically investigated the rich diversity of $\mathcal{PT}$-symmetry breaking and restoring transitions that arise in a two-dimensional, continuum, circular domain with uniform gain or loss potentials confined to parity-symmetric disk-shaped regions. The resulting two-dimensional geometry lacks any continuous symmetry and the hope of any effective, dimensional reduction. Therefore, using a interface-conforming discretization, we have numerically solved the resultant generalized eigenvalue problem for lowest few eigenvalues. We have found multiple $\mathcal{PT}$-symmetry breaking and restoring transitions, that occur generically as the gain-loss strength $\gamma$ is increased while other system parameters are fixed. We have also found that while the threshold $\gamma_\mathrm{PT}$ scales inversely with the size of the gain-loss regions, it shows a non-monotonic dependence on the separation $d$, with a marked increase in $\gamma_\mathrm{PT}$ that occurs as the gain-loss domains approach the boundary of the fiber. Our results show that stable, propagating modes are supported in multi-core fiber with gain and loss regions. The difference in the nominal size of the modes and size of gain-loss cores, due to the absence of index contrast, is primarily responsible for the non-trivial dependence of $\gamma_\mathrm{PT}$ on the location of gain and loss domains. 

The lattice and continuum $\mathcal{PT}$-symmetric models, particularly those relevant in optics, have focused on two categories. For models in the first category, the gain regions span half the domain, with the loss-region spanning the remaining, parity-symmetric counterpart~\cite{Znojil2001,Ruter2010,Regensburger2012,Feng2017,Miri2019,Ozdemir2019}. For models in the second category, the gain-loss regions are highly localized (measure zero) relative to the size of the domain~\cite{Joglekar2010,Jin2009,Harter2016}. In each case, increasing the gain-loss strength in the $\mathcal{PT}$-broken region leads to more unstable modes. On the other hand, our single-parameter model shows that when gain-loss regions occupy a finite, tunable fraction of the domain, multiple $\mathcal{PT}$-breaking transitions are possible. The emergence of stable, propagating modes with real eigenvalues in response to {\it increasing gain-loss strength} means that such fibers can serve to support both amplifying and propagating modes. These results suggest that significant threshold engineering can be carried out by using spatially distributed gain and loss domains in a bounded region with no symmetries beyond the discrete, $\mathcal{PT}$ symmetry. 


\section*{Acknowledgements}
T. Gratcheva gratefully acknowledges an NSF RTG Undergraduate Summer Fellowship (NSF Grant 2136228). This work was also supported in part by AFOSR grant FA9550-23-1-0103 (JG), NSF Grant 2245077 (TG, JG) and ONR Grant No. N00014-21-1-2630 (YJ). 

\bibliography{ptyj1,ptyj2,numer}

\end{document}